\documentclass[11pt,a4paper]{article}

\usepackage[utf8]{inputenc}
\usepackage[margin=2.5cm]{geometry}
\usepackage{setspace}
\usepackage{graphicx}
\usepackage{amsmath,amssymb,bm}
\usepackage{xcolor}
\usepackage{caption}
\usepackage{siunitx}

\usepackage{booktabs}
\usepackage{makecell}
\usepackage{array}
\usepackage{tabularx}
\usepackage{float}
\usepackage{subcaption}

\usepackage[backend=biber,sorting=none]{biblatex}
\graphicspath{{figures/}}

\definecolor{markerorange}{RGB}{210,105,0}
\definecolor{markerred}{RGB}{180,20,20}
\definecolor{markerblue}{RGB}{0,90,170}
\definecolor{markergreen}{RGB}{0,125,80}

\newif\ifshowdraftmarkers \showdraftmarkerstrue

\usepackage{tikz}
\usetikzlibrary{shapes.geometric, arrows, positioning}
\tikzstyle{startstop} = [rectangle, rounded
corners, minimum width=3cm, minimum height=1cm,text centered, draw=black, fill=blue!10]
\tikzstyle{process} =
[rectangle, minimum width=3.5cm, minimum height=1cm, text centered, text width=5.5cm, draw=black, fill=orange!10]
\tikzstyle{io} = [trapezium, trapezium left angle=70, trapezium right angle=110, minimum width=3cm, minimum
height=1cm, text centered, draw=black, fill=green!10]
\tikzstyle{arrow} = [thick,->,>=stealth]

\usepackage[
    colorlinks=true, citecolor=black, linkcolor=black, urlcolor=black, filecolor=black
]{hyperref}
\usepackage{xr-hyper}
 
\usepackage{verbatim}

\usepackage{titlesec}

\titleformat{\section}
  {\normalfont\fontsize{13.5}{16}\selectfont\bfseries}
  {\thesection}
  {1em}
  {}
  
\begin{document}

\begin{center}

{\Large\bfseries
Portable Vector NV-Diamond Magnetometer for Shot-Noise-Limited, Drift-Free Operation in Unshielded Environments
\par}

\vspace{0.3cm}

{\large
	Annirudh K P$^{1,2,\dagger}$, Vinayak Rane$^{2,\dagger}$, Shradha Atakar$^{1}$, Sanika Joshi$^{1}$,
	Maheshwar Mangat$^{1}$, Jay Gharat$^{2}$, Siddharth Tallur$^{1,*}$, Kasturi Saha$^{1,2,*}$ \par}

\vspace{0.2cm}

{\small
	$^{1}$Department of Electrical Engineering, Indian Institute of Technology Bombay, Mumbai 400076, Maharashtra, India \\ \quad $^{2}$Qmet Tech Foundation, Mumbai 400076, Maharashtra, India \par}

{\footnotesize
	$^{\dagger}$These authors contributed equally.	\quad $^{*}$Corresponding authors.  \par}

\vspace{0.12cm}

\begin{abstract}	
    Ensemble nitrogen-vacancy (NV) diamond magnetometers combine high sensitivity with vector-field reconstruction, but practical deployment is limited by errors arising from high-frequency laser noise during short-duration operations and slow-varying gain fluctuations and offset drift during long-term operation. Here, we present an integrated digital architecture for achieving NV magnetometry stability across distinct timescales. A dynamic differential readout continuously balances fluorescence and reference channels to suppress correlated laser noise. Second-derivative Lorentzian lineshape tracking enables in-situ correction of slope variations arising from slow changes in gain and optical excitation. We further identify temperature-induced bias-magnet fluctuations as a dominant source of long-term drift and introduce a magnetic eigenvector transformation that uses the intrinsic response of NV resonances to eliminate these variations. We show, under unshielded ambient conditions, that this architecture achieves an off-resonance noise factor of 1.0 $\pm$ 0.1, matching the fundamental limit with ten-fold suppression of long-term drift, enabling stable, field-ready quantum magnetometry.
\end{abstract}

\end{center}

\vspace{0.4cm}

\label{sec:introduction}

Ensemble nitrogen-vacancy (NV) centres in diamond represent a compelling solid-state platform for high-sensitivity quantum magnetometry owing to their three-axis sensing capabilities, room-temperature operation, and long spin coherence times~\cite{Barry2020RMP}. Consequently, these systems have rapidly transitioned towards fiber-coupled and portable instruments targeting applications such as geomagnetic surveying~\cite{Graham2025Road,Halde2025Field}, biomagnetic diagnostics, and non-destructive industrial testing, among others~\cite{Webb2019,Patel2020,Graham2023,Zhang2022Fiber,Kim2019,Sekiguchi2024,Kumar2024,Huang2024,Zheng2024,Liu2024Fiber,Wang2025Integrated,Liu2026Chip,Xie2024MEMS}.
Prior studies~\cite{Barry2020RMP,Wang2025Laser,Acosta2010} have identified two dominant short-term noise sources in such systems:
laser relative intensity noise (RIN) and rapid thermal shifts of the resonance frequencies arising from the temperature dependence of the zero-field splitting 
$(D_{ZFS})$. High-sensitivity diamond magnetometry reports in literature have since demonstrated suppression of these local fluctuations with considerable success; rapid optical intensity fluctuations are routinely canceled through hardware-based common-mode balanced photodetection~\cite{Sturner2021}, while short-term thermal line wandering is compensated by continuously tracking Zeeman-split transition pairs along a selected crystallographic axis and utilizing their differential frequency response for magnetometry~\cite{Clevenson2018}.

However, when transitioning from controlled laboratory systems to unshielded, long-term, real-world operation, this conventional stabilisation paradigm becomes fundamentally inadequate. Extended multi-hour measurements introduce two
bottlenecks that conventional hardware frameworks cannot fully resolve. First, slow, time-dependent variations in laser power alter the effective gains of the signal and reference paths. The resulting gain imbalance progressively degrades the common-mode rejection ratio, allowing common-mode, high-frequency laser noise to leak through the demodulation stage. This vulnerability is evident across state-of-the-art systems as a persistent \(6\text{--}20\,\text{dB}\) elevation of the magnetically insensitive off-resonance amplitude spectral density (ASD) above the fundamental electronic dark-noise floor~\cite{Schloss2018,Kumar2024,Wang2026ClosedLoop}. 
Active hardware architectures have been developed to mitigate this path imbalance~\cite{Wang2026ClosedLoop}; nevertheless, complete and sustained laser-noise cancellation over long measurement durations remains difficult to achieve.

Second, long-term offset stability remains a significant challenge in unshielded NV magnetometers. We demonstrate that the widely trusted, conventional Zeeman-split pair differential tracking strategy fails to eliminate this slow thermal drift that manifests over multi-hour operation, leaving the absolute sensor offset entirely vulnerable to ambient thermal variations. 
A recent field-deployed implementation uses an auxiliary temperature sensor placed near the magnet assembly together with a calibrated temperature-to-field correction~\cite{Halde2025Field}. The accuracy of such compensation is then set not only by the thermometer itself, but also by how faithfully its measured temperature changes represent those of the magnetic field generated by the assembly. This becomes non-trivial even for high-stability samarium-cobalt (\(\mathrm{SmCo}\)) magnets, whose field can vary by approximately \(40\,\text{nT}/0.1\,\text{K}\)~\cite{ConstantinidesRTC}. The added sensor also introduces packaging and calibration requirements that become increasingly restrictive in high-sensitive, compact instruments. Consequently, optimizing short-term precision and ensuring long-term, drift-free stability have historically existed as mutually exclusive instrumentation design paradigms.

In this work, we present, for the first time, a comprehensive architectural solution to address these limitations through a hardware-agnostic quantum sensing framework that addresses technical noise and environmental drift across distinct processing timescales. We implement a time-adaptive digital differential scheme in which the relative gain between the fluorescence and optical-reference channels is continuously tracked, enabling broadband laser-noise suppression without modification of the physical optical path. We then introduce a real-time discriminator-slope correction engine that exploits the second derivative of the optically detected magnetic resonance (ODMR) lineshape to 
compensate variations in optical gain and resonance contrast in real-time. 
Additionally, we establish the theoretical and mathematical framework for a novel drift-removal strategy termed magnetic eigenvector transformation. This formalism uses the diamond's own resonance frequencies as an intrinsic, self-contained drift probe, allowing slow common-mode variations originating from the bias magnets to be isolated from genuine external magnetic-field signatures. Finally, we validate the complete framework through continuous measurements under uncontrolled and unshielded environments. By combining strong suppression of the laser-noise floor with elimination of long-term magnetic baseline drift, the resulting magnetometer enables robust, autonomous operation in unshielded real-world environments.

\section{Optical Readout Geometry for Common-Mode Rejection} \label{sec:HD}

\begin{figure}[t]
	\centering \includegraphics[width=\textwidth]{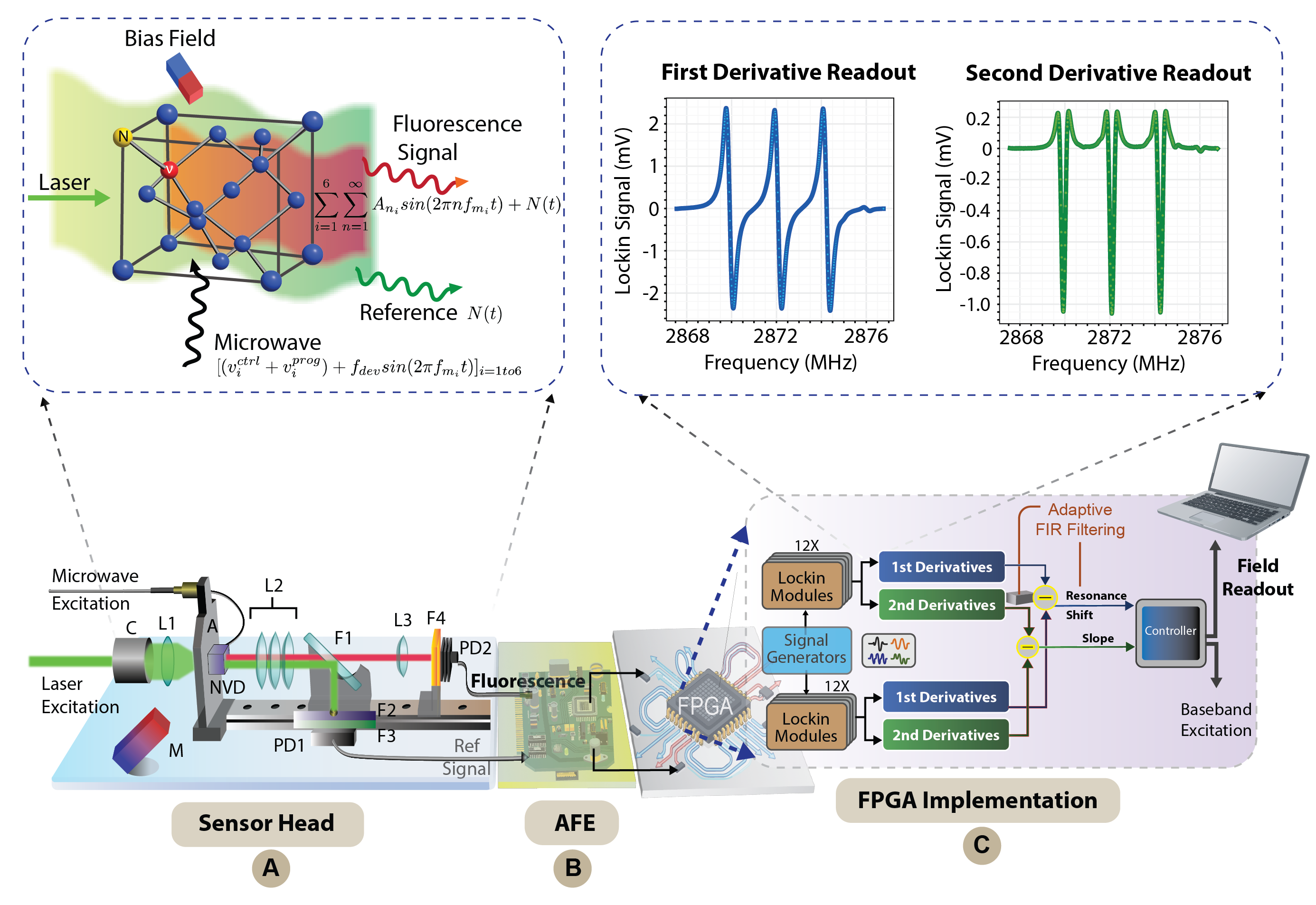}

	\caption{\textbf{Setup Schematic: Post-diamond differential optical readout and derivative-resolved ODMR tracking.} The nitrogen-vacancy (NV) ensemble is optically pumped under a static bias field and interrogated using six frequency-modulated microwave tones addressing six ODMR transitions. NV fluorescence provides the magnetic signal, while residual transmitted green pump light serves as a correlated laser-noise reference. The pump is focused through the nitrogen-vacancy diamond (NVD), and the transmitted pump and fluorescence are collected along the same path and separated after the diamond by dichroic element F1. The fluorescence is detected by PD2 and the residual pump by PD1. The resonance frequency is identified from the point of maximum local ODMR slope, establishing the frequency-to-detuning calibration. The local curvature at this point tracks variations in the first-derivative discriminator slope. After the analog front end (AFE), both signals are digitised and processed in a field-programmable gate array (FPGA). Twelve parallel lock-in channels per detector recover the first- and second-derivative responses of the six transitions. An adaptive finite-impulse-response (FIR) filter matches the reference to the fluorescence channel for digital laser-noise cancellation. The corrected first derivative determines the resonance detuning, while the second derivative provides slope correction for the digital frequency-lock controller. Here: C:fiber collimator; L1-L3, lenses; M, bias magnet; A, microwave antenna; F1-F4, optical filters; PD1-PD2, photodetectors.}
	\label{fig:system_architecture}
\end{figure}

Maximizing the short-term sensitivity of NV magnetometers requires strong suppression of laser relative-intensity noise (RIN). A conventional approach is to split a small fraction of the excitation beam before the diamond by typically using a $90:10$ or $99:1$ fused-fiber coupler and then use this branch as the reference for balanced detection~\cite{Wang2026ClosedLoop, Wang2025Laser}. This reference samples the laser before the sensing path (pre-diamond) and therefore does not experience the same path-dependent optical variations as the fluorescence channel. Internal reflections, scattering, and path-dependent losses introduced through the diamond and collection optics can modify the amplitude and phase of the detected laser noise. These contributions are absent from a pre-diamond reference, leaving a residual mismatch that limits common-mode rejection. To reduce this mismatch, we
perform the fluorescence--reference separation entirely after the diamond [Fig.~\ref{fig:system_architecture}]. The full green excitation beam is collimated by \(\mathrm{C}\), focused by \(\mathrm{L1}\), and passed through the nitrogen-vacancy diamond (\(\mathrm{NVD}\)). The transmitted pump and generated NV fluorescence are collected together by \(\mathrm{L2}\) and co-propagated up to the dichroic element \(\mathrm{F1}\). At \(\mathrm{F1}\), the fluorescence is routed through \(\mathrm{L3}\) and \(\mathrm{F4}\) onto \(\mathrm{PD2}\), thereby capturing the fluorescence, while the residual green pump is directed through \(\mathrm{F2}\) and \(\mathrm{F3}\) onto \(\mathrm{PD1}\), forming the optical reference.

The reference is therefore sampled only after traversing the sensing path. Laser fluctuations introduced before and through the diamond remain correlated between the two channels, while the NV fluorescence retains the microwave-dependent ODMR response. This preserves the shared optical perturbations through the sensor head while confining the residual mismatch largely to the post-separation detection paths, where it can be corrected digitally. The remaining mismatch, including detector gain and filtering, is corrected using the adaptive Finite Impulse Response (FIR) filter. A direct experimental comparison of the noise floors obtained with pre- and post-diamond reference collection is provided in the Supplementary Information. This post-diamond differential measurement scheme is used for all subsequent measurements.

\section{Three-axis Readout with Digital Noise Cancellation}

\begin{figure*}[!t]

	\centering

	\includegraphics[width=\textwidth]{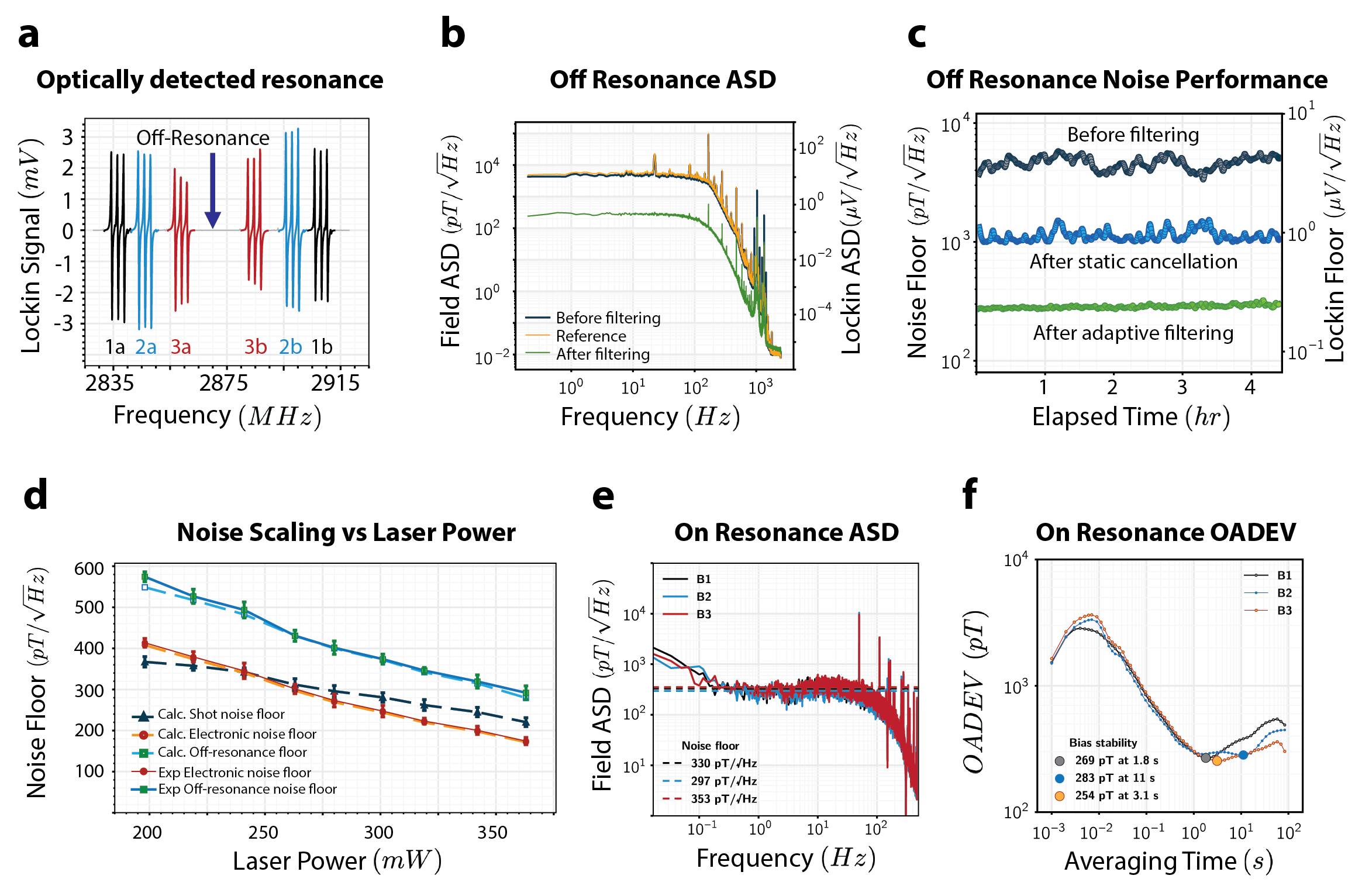}

	\caption{\textbf{Three-axis readout and adaptive laser-noise cancellation.} \textbf{(a)} Demodulated ODMR spectrum containing six addressed transitions, grouped into three Zeeman-split pairs used to reconstruct \(B_1\), \(B_2\), and \(B_3\). The marked off-resonance point is used to measure the readout-noise floor without ODMR frequency-to-signal transduction.  \textbf{(b)} Off-resonance amplitude spectral densities (ASDs) of the fluorescence, optical-reference, and adaptively corrected channels. The signals are acquired at \(5~\mathrm{kS\,s^{-1}}\) with lock-in amplifier filter set as eighth-order filter with low-pass cut-off frequency of approximately \(102~\mathrm{Hz}\). Matching the reference with the adaptive FIR removes the correlated laser-noise component and reduces the residual towards the electronic noise floor.  \textbf{(c)} Sliding \(1\)--\(100~\mathrm{Hz}\) noise density over \(\approx 4.5~\mathrm{hrs}\). The residual from a fixed subtraction increases as the relative response of the two channels changes, whereas the adaptive filter maintains the channel balance throughout the acquisition.  \textbf{(d)} Calculated and measured noise floors as the optical excitation is varied from approximately \(200~\mathrm{mW}\) to \(363~\mathrm{mW}\). The corrected off-resonance measurements follow the calculated total noise floor due to electronic noise and photon shot-noise across the power range.  \textbf{(e)} On-resonance ASDs of the simultaneously reconstructed \(B_1\), \(B_2\), and \(B_3\) projections. The narrow feature at \(50~\mathrm{Hz}\) originates from mains magnetic pickup.  \textbf{(f)} Overlapping Allan deviation (OADEV) obtained from \(10\)-minute continuous three-axis measurements. After the initial response of the lock-in and tracking chain, the traces follow the \(\tau^{-1/2}\) dependence expected for white noise before slow drift becomes visible at longer averaging times.}

	\label{fig:adaptive_readout}

\end{figure*}

The three-axis measurement uses six simultaneously addressed ODMR transitions, grouped into three Zeeman-split pairs associated with three crystallographic NV orientations ~\cite{Schloss2018,Clevenson2018,Zhao2019} [Fig.~\ref{fig:adaptive_readout}(a)]. The frequency difference within each pair gives one magnetic projection, \(B_1\), \(B_2\), or \(B_3\), while shifts common to both splits are rejected. This suppresses common-mode changes of the ODMR spectrum, such as that produced by changes in the $D_{ZFS}$, while retaining the differential response required for magnetic-field reconstruction. The three projections are acquired simultaneously through the same optical, microwave, and digital readout chain, so that three-axis reconstruction does not require independent sensing hardware for each axis. The off-resonance point marked in Fig.~\ref{fig:adaptive_readout}(a) is used to characterize the readout independently of the magnetic response. With the microwave carriers 
detuned far
from the ODMR lines, fluctuations in microwave frequency are no longer converted into a fluorescence signal, whereas only the optical and electronic noise of the detection chain remain. The residual noise measured off-resonance therefore provides a direct measure of the performance of the optical common-mode rejection.

\label{sec:adaptive_readout}

The fluorescence and reference channels carry strongly correlated laser-intensity noise, but they are not identical
copies of one another. Their relative gain and delay depend on the optical filters, photodetectors, analog front ends, and other differences in experimental realization of the two optical paths. These quantities also change slowly during operation, so a fixed subtraction cannot maintain the optimum balance indefinitely ~\cite{Wang2025Laser,Wang2026ClosedLoop}. We instead match the reference to the fluorescence channel digitally after lock-in demodulation using a centered FIR filter. An initial set of coefficients is obtained from an off-resonance calibration using regularized least-squares fitting. During acquisition, the coefficients are updated with a bounded normalized least-mean-square (NLMS) algorithm~\cite{Slock1993}. The bounds are set around the calibrated solution so that the adaptive filter can follow slow changes in the relative response of the two optical channels without drifting arbitrarily from the calibrated state.

The spectral comparison in Fig.~\ref{fig:adaptive_readout}(b) suggests that most of the broadband structure present in the fluorescence channel is also present in the optical reference. Once the filtering is performed, the correlated component is suppressed by more than an order of magnitude across the \(1\)--\(100~\mathrm{Hz}\) measurement band, bringing the residual close to the detector-noise floor. The noise performance with time is shown in Figure~\ref{fig:adaptive_readout}(c) comparing the off-resonance noise floor before filtering, after static cancellation, and after adaptive filtering over more than four hours. Static cancellation provides only partial common-mode suppression and remains sub-optimal compared with adaptive filtering. Its performance also degrades as the relative response of the two optical channels drifts with time. By continuously updating the FIR coefficients, the adaptive filter maintains a lower and more stable residual throughout the acquisition.

\begin{table}[!t]
    \centering

    \caption{\textbf{Calculated readout-noise limits and experimentally measured three-axis noise floors.} All values are referred to the differential Zeeman-split pair magnetic readout. The calculated off-resonance floor is the quadrature sum of the electronic and photon shot-noise contributions, with the residual laser contribution negligible after cancellation. Experimental off-resonance values are measured after adaptive cancellation over the \(1\)--\(100~\mathrm{Hz}\) band. The \(B_1\), \(B_2\), and \(B_3\) values correspond to simultaneous six-transition three-axis operation in an unshielded environment.}
    \label{tab:noise_budget}

    \renewcommand{\arraystretch}{1.16}

    {\footnotesize
    \begin{tabular}{@{}l r@{\hspace{1.5em}}l r@{}}
        \toprule

        \textbf{Noise metric}
        & \textbf{Field noise}
        & \textbf{Noise metric}
        & \textbf{Field noise}
        \\

        & \textbf{(pT/\(\sqrt{\mathrm{Hz}}\))}
        &
        & \textbf{(pT/\(\sqrt{\mathrm{Hz}}\))}
        \\

        \cmidrule(lr){1-2}
        \cmidrule(lr){3-4}

        \multicolumn{2}{c}{\textit{Calculated readout limits}}
        &
        \multicolumn{2}{c}{\textit{Experimentally verified limits}}
        \\

        Electronic noise floor
        & \(174 \pm 6\)
        & Electronic noise floor
        & \(177 \pm 4\)
        \\

        Photon shot noise floor
        & \(225 \pm 5\)
        & Off-resonance noise floor
        & \(288 \pm 11\)
        \\

        Residual laser noise floor
        & \(\approx 0\)
        & \(B_1\) white-noise floor
        & \(330 \pm 23\)
        \\

        Off-resonance noise floor
        & \(284 \pm 8\)
        & \(B_2\) white-noise floor
        & \(296 \pm 26\)
        \\

        &
        &
        \(B_3\) white-noise floor
        & \(353 \pm 21\)
        \\

        \bottomrule
    \end{tabular}
    }
\end{table}

The comparison with the calculated noise budget in Table~\ref{tab:noise_budget} places this suppression on an absolute scale. After adaptive cancellation, the measured off-resonance floor converges directly with the combined limit imposed by electronic noise and photon shot-noise sources, leaving no resolvable laser-noise-dominated excess within the measured band. The optical-power sweep in Fig.~\ref{fig:adaptive_readout}(d) confirms the same result over a wider operating range. As the excitation power changes, the corrected measurements continue to follow the calculated readout floor rather than developing a separate laser-noise-limited plateau. The cancellation is therefore maintained as the optical operating point is varied, rather than being restricted to a single calibrated power.
Finally, the same processing is used during full three-axis operation, with all six microwave carriers tracking their respective ODMR transitions. The three reconstructed magnetic projections retain comparable short-term noise floors and show the expected white-noise behaviour in the Allan deviation analysis [Fig.~\ref{fig:adaptive_readout}(e,f)]. Importantly, the adaptive filter acts on the component correlated with the optical reference, not on the magnetic response itself (see Supplementary Information). Magnetic features are therefore retained in the reconstructed field channels; this is also seen in the \(50~\mathrm{Hz}\) line-frequency magnetic pickup in Fig.~\ref{fig:adaptive_readout}(e). The absence of this feature under off-resonance operation rules out direct electrical pickup, supporting its magnetic origin.

\section{ODMR Second-Derivative-Stabilized Digital Frequency Locking}

\label{sec:second_derivative}

The first-derivative of the ODMR signal provides the operating point required for resonance tracking, but its conversion from lock-in voltage to frequency detuning depends on the local ODMR slope. This slope changes with the optical operating point: variations in laser power modify the fluorescence level, ODMR contrast, and linewidth, and therefore change the magnitude of the first-derivative signal of the ODMR produced by the same frequency offset~\cite{Lei2025}. A fixed calibration of the time-dependent voltage-to-frequency conversion yields an effective loop gain that drifts during operation. We correct this effect using the second derivative of the ODMR lineshape. Under frequency modulation, the first-derivative channel measures the local detuning about resonance, while the second-derivative channel measures the ODMR curvature~\cite{Wojciechowski2019}. At the resonance point, this curvature is directly related to the local slope of the first-derivative response (see Methods, Sec.~\ref{sec:methods_tracking}). Both channels are therefore acquired simultaneously for each of the six addressed transitions, and the second-derivative response is calibrated against the measured ODMR lineshape, modulation depth, and demodulation phase. During operation, it provides the instantaneous scaling required to convert the first-derivative error into frequency detuning. The discriminator slope is therefore tracked in situ during closed-loop operation, without requiring repeated ODMR calibration sweeps.

Since drift is a slow phenomenon, the second-derivative signal is filtered with a low-pass filter before normalization so that high frequency noise in the second-derivative channel does not result in rapid modulation of the loop gain. The resulting slope-corrected detuning is then passed to an \(\alpha\text{--}\beta\) tracker, which estimates the resonance point frequency and its drift rate and updates the microwave carrier frequency accordingly. The centre-frequency and drift-rate estimates predict the subsequent resonance position, while the measured detuning corrects this prediction and updates the drift rate. The controller therefore acts on a frequency-domain error rather than directly on the raw lock-in voltage, while slow changes in ODMR contrast and linewidth are absorbed into the continuously updated slope estimate.

\begin{figure}[!t]

	\centering

	\includegraphics[width=0.8\linewidth]{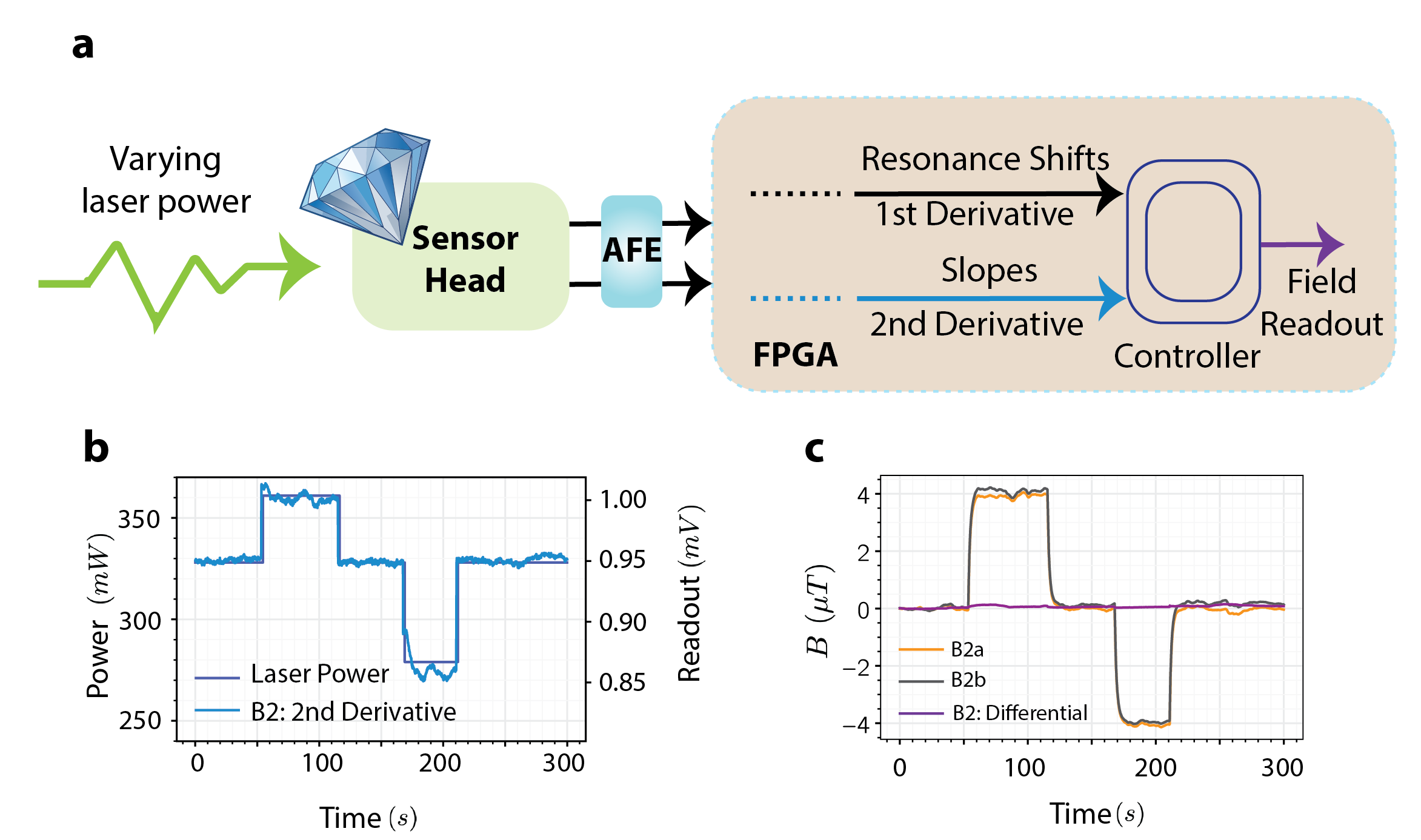}

	\caption{
    \textbf{Second-derivative normalization of the digital frequency lock.}
		\textbf{(a)} Control architecture. The first-derivative ODMR channel provides the resonance-detuning error, while the second-derivative channel provides the corresponding discriminator-slope estimate used to normalize the error before the frequency controller.  \textbf{(b)} Applied laser-power sequence and measured second-derivative response of the \(B_2\) resonance pair. The curvature follows the optical-power variation and tracks the corresponding change in discriminator slope. \textbf{(c)} Simultaneously tracked \(B_{2a}\) and \(B_{2b}\) resonances and their differential magnetic readout. Laser-induced heating shifts both resonances in common through the temperature dependence of the zero-field splitting, while the differential Zeeman readout suppresses this common-mode contribution.
}

	\label{fig:second_derivative}

\end{figure}

The frequency-lock bandwidth is set to approximately \(8~\mathrm{Hz}\), well below the \(\approx102~\mathrm{Hz}\) lock-in cut-off frequency. This bandwidth applies only to the carrier-tracking loop and does not limit the magnetic measurement itself~\cite{Wang2023HDR}. Slow changes in the ODMR centre are tracked by the controller, while faster magnetic variations remain in the slope-normalized residual detuning. The measured resonance frequency is reconstructed from the controller correction together with this residual term, preserving the full DC--\(100~\mathrm{Hz}\) readout bandwidth. The controlled laser-power sweep illustrated in Fig.~\ref{fig:second_derivative}(a) tests the slope normalization and resonance tracking simultaneously.

As the optical power changes, the second-derivative channel follows the change in ODMR curvature [Fig.~\ref{fig:second_derivative}(b)], while \(B_{2a}\) and \(B_{2b}\), the two Zeeman splits of the \(B_2\) projection, shift together because of laser-induced heating [Fig.~\ref{fig:second_derivative}(c)]. The same paired-branch arrangement applies to the other two measurement axes; the \(B_2\) pair is shown here as a representative example.
The pairwise difference removes the common thermal shift, while the second-derivative normalization corrects the accompanying change in the slope. The result is a frequency lock whose response remains calibrated even as the ODMR lineshape changes during operation.

\section{Long-term Stabilization and Drift Correction via Magnetic Eigenvector Transformation}

\label{sec:SD}

The conventional Zeeman-split pair differential sensing compensates temperature-induced shifts of the \(D_{\mathrm{ZFS}}(T)\) parameter~\cite{Xie2023DoubleTransition}, but it cannot correct errors due to temperature-induced changes in the bias field itself~\cite{Childress2025BiasFree}. 
For a Zeeman-split pair, a magnetic-field change produces equal and opposite shifts of the two resonance frequencies and therefore appears in the differential coordinate, whereas a change in the zero-field splitting produces a common shift of both resonances and appears in the common-mode coordinate. These coordinates would ideally remain independent. However, in a temperature-unregulated operation, the ambient temperature couples them through the permanent magnets used for applying a bias magnetic field: the same temperature change that shifts the \(D_{\mathrm{ZFS}}\) also changes the magnet remanence, producing a real magnetic field shift. Consequently, this ambient thermal perturbation stamps an identical thermodynamic signature onto both the common-mode and differential channels concurrently. Rather than remaining isolated, the parasitic bias-field drift corrupts the active magnetic readout directly, manifesting as slow, multi-hour baseline drifts that track the environmental thermal profile.
 
\begin{figure}[!t]

	\centering

	\includegraphics[width=\textwidth]{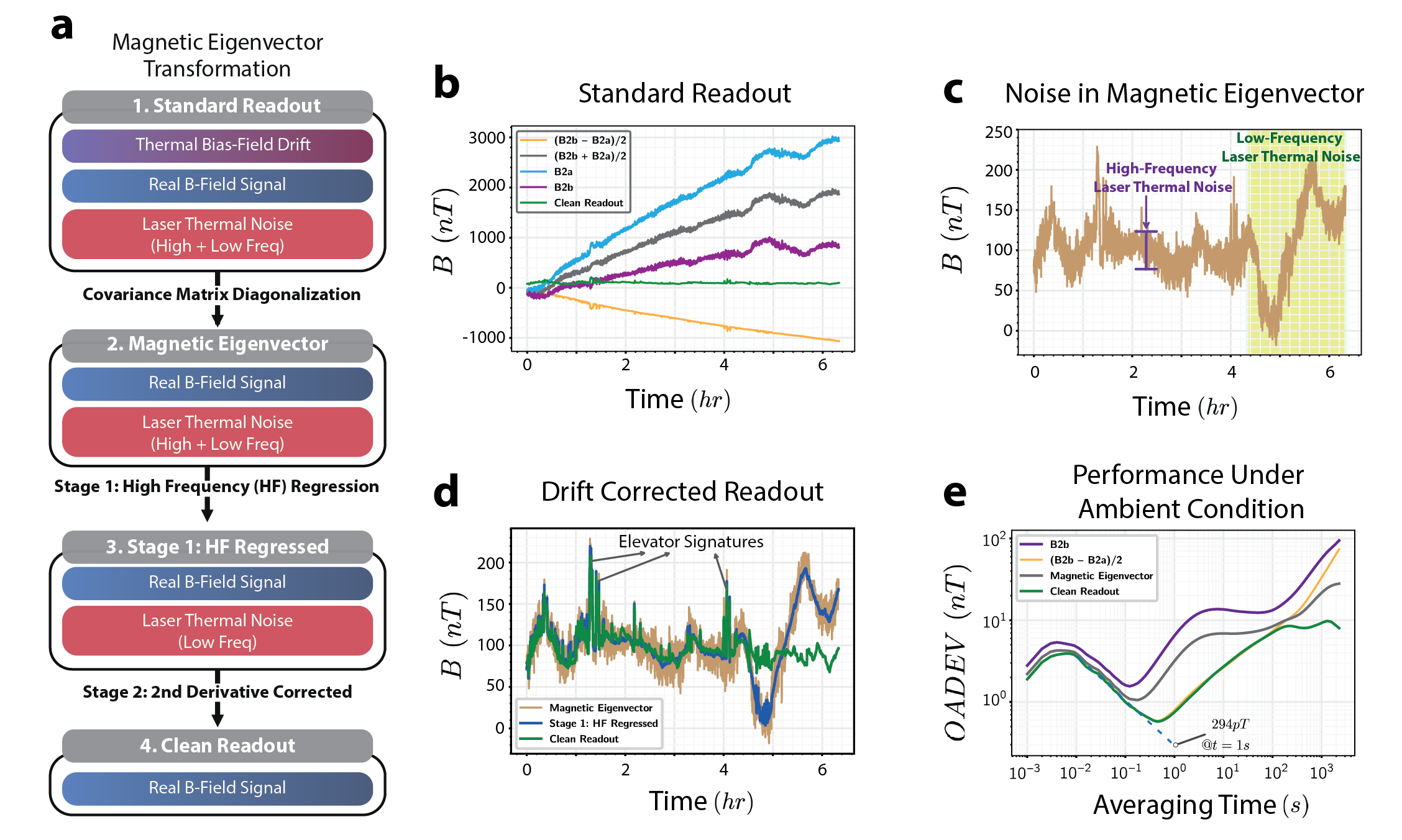}

	\caption{\textbf{Long-term drift removal using magnetic eigenvector transformation.}
		\textbf{(a)} Processing sequence. Covariance-matrix diagonalisation first isolates the temperature-correlated bias-field drift from the magnetic coordinate. High-frequency (HF) regression then removes the remaining fast laser-correlated component, followed by second-derivative correction of the slower laser-induced drift.  \textbf{(b)} Six-hour long \(B_2\) resonance measurement acquired in an unshielded laboratory with active temperature regulation disabled. The \(B_{2a}\) and \(B_{2b}\) transitions exhibit strong common-mode thermal variation, while the conventional differential magnetic coordinate develops a pronounced drift that is anti-correlated with the common-mode thermal coordinate. The ambient temperature varied by approximately \(22\,^{\circ}\mathrm{C}\text{--}26\,^{\circ}\mathrm{C}\). \textbf{(c)} Magnetic eigenvector obtained after covariance diagonalization. The dominant ambient bias-field drift is suppressed, leaving the magnetic signal together with residual high-frequency laser noise and a slower laser-induced component.  \textbf{(d)} Successive correction of the magnetic eigenvector. HF regression removes the fast laser-correlated contribution, while the second-derivative correction removes the remaining slow optical drift. Short magnetic transients, including the marked elevator signatures, are retained.  \textbf{(e)} Overlapping Allan deviation (OADEV) of the single \(B_{2b}\) transition, conventional differential readout, raw magnetic eigenvector, and final corrected eigenvector. The corrected channel retains a short-term OADEV of approximately \(294\,\mathrm{pT}\) at \(\tau=1\,\mathrm{s}\), i.e., \(294~\mathrm{pT/\sqrt{Hz}}\) noise density, while reducing the long-term drift by approximately one order of magnitude.}

	\label{fig:qmet}

\end{figure}

To eliminate these coupled errors, we introduce a software-defined magnetic eigenvector transformation framework, whose multi-timescale digital processing sequence is outlined in Fig.~\ref{fig:qmet}(a). Rather than relying on an external thermometer and a separate temperature-to-field hardware calibration matrix, this framework directly exploits the joint thermodynamic correlation stamped across both channels by applying an online covariance diagonalization to the co-evolving magnetic and thermal coordinates. Because the parasitic thermal drift vector and genuine external magnetic targets possess linearly independent projection vectors within this shared measurement space, the eigen-decomposition decouples them seamlessly. The dominant temperature-correlated variance is isolated within a primary eigen-coordinate, leaving the orthogonal secondary eigen-coordinate to project a pure magnetic trace with strongly reduced bias-field drift.

A continuous six-hour measurement (Fig.~\ref{fig:qmet}(b)) demonstrates the limitations of conventional Zeeman-split pair differencing.
With the laboratory temperature control disabled, the differential coordinate drifts by approximately 1000 nT,
even though the large common-mode thermal shifts of the two transitions are rejected. The differential magnetic coordinate is strongly anti-correlated with the common-mode thermal coordinate, identifying temperature-dependent bias-magnet strength as the dominant source of the drift. This direct correlation thereby establishes an ideal testbed for eigenvalue diagonalization.

Covariance diagonalization first yields the raw magnetic eigen-coordinate (Fig.~\ref{fig:qmet}(c); Methods, Sec.~\ref{sec:methods_eigen}), strongly suppressing the long-term ambient bias-field drift. However, the resulting magnetic eigenvector still carries an unwanted projection of the laser-induced thermal noise, which is systematically removed in two sequential stages. The high-frequency noise is addressed first by extracting its profile from the common-mode thermal coordinate and executing a linear regression against the raw magnetic eigenvector (see Supplementary).  The remaining slow component is subsequently removed using the baseline-subtracted, low-pass-filtered second-derivative signal, \(S_{2d,\mathrm{LP}}(t)\), as an uncorrupted proxy for slow laser-power variation (see Supplementary). These complementary digital blocks operate on distinct timescales and leave macroscopic magnetic transients intact, as demonstrated by the clear capture of building elevator displacement signatures in Fig.~\ref{fig:qmet}(d).

The OADEV results reported in Fig.~\ref{fig:qmet}(e) illustrate the effect of this scheme on long-term drift. Conventional Zeeman differencing performs well at short averaging times but degrades as temperature-driven changes in the bias field accumulate. At \(\tau = 2\times10^{3}~\mathrm{s}\), the OADEV rises to approximately \(80~\mathrm{nT}\), whereas the fully corrected magnetic eigenvector remains near \(8~\mathrm{nT}\). The correction therefore improves the long-term stability by roughly an order of magnitude without sacrificing the short-term magnetic response. We further tested the method in a $12-$hour open-environment measurement spanning approximately \(28\text{--}32\,^{\circ}\mathrm{C}\) (see Supplementary).
The larger temperature excursion produced substantial laser mode and power instability, preventing reliable use of the secondary second-derivative correction. The primary eigenvector transformation nevertheless continued to suppress the dominant temperature-induced bias-field drift.

The magnitude of this effect remains relevant even for high-stability permanent magnets. The reversible temperature coefficient of remanence, \(\alpha_{B_r}\), is typically \(\sim -(0.10\text{--}0.12)\,\%\,^{\circ}\mathrm{C}^{-1}\) for sintered NdFeB and \(\sim -0.035\,\%\,^{\circ}\mathrm{C}^{-1}\) for conventional \(\mathrm{Sm_2Co_{17}}\) magnets~\cite{ConstantinidesRTC}. Even temperature-compensated SmCo grades retain coefficients of order \(10~\mathrm{ppm\,K^{-1}}\)~\cite{EEC_SmCoTC,Liu2019SmCo}. For a \(1~\mathrm{mT}\) bias field, a \(0.1~\mathrm{K}\) temperature change corresponds to approximately \(100\text{--}120~\mathrm{nT}\) for NdFeB, \(35~\mathrm{nT}\) for conventional SmCo, and about \(1~\mathrm{nT}\) even for compensated SmCo. These shifts are still significant for sub-nanotesla magnetometry. While external thermometry can be deployed for hardware-level correction, it introduces restrictive packaging constraints, requires tedious temperature-to-field calibrations, and remains fundamentally limited by thermal lag between the thermometer and the magnet.  

The software-defined eigenvector approach completely bypasses these physical hardware limits. By dynamically estimating the temperature-correlated bias-field drift directly from the co-evolving NV resonance coordinates themselves, our framework permanently decouples environmental thermal fluctuations from the true magnetic target in software, establishing a self-contained, calibration-free operational paradigm for unshielded outdoor magnetometry.

\section{Performance Benchmark} \label{sec:benchmark}

Absolute sensitivity alone is a poor measure of how effectively an NV magnetometer suppresses technical noise, since the reported field sensitivity also depends on diamond volume, optical power, collection efficiency, ODMR contrast, linewidth, and detector gain. To compare the readout architectures on a common basis, we define the noise factor: \[
	\mathcal{F} = \frac{S}{S_{\mathrm{limit}}}, \qquad S_{\mathrm{limit}} =
	\sqrt{S_{\mathrm{elec}}^{2}+S_{\mathrm{PSN}}^{2}},
\] where \(S\) is the measured field-referred noise density and \(S_{\mathrm{limit}}\) is the corresponding readout
limit set by the electronic and photon shot-noise contributions (see Methods, ``Fundamental readout-noise evaluation,'' and Supplementary Information).

We evaluate \(\mathcal{F}\) separately for on-resonance ODMR operation (magnetically sensitive regime), \(\mathcal{F}_{\mathrm{on}}\), and under magnetically insensitive off-resonance conditions, \(\mathcal{F}_{\mathrm{off}}\). Thus, \(\mathcal{F}=1\) denotes operation at the calculated fundamental readout limit.

\begin{table}[!t]
	\centering \small \setlength{\tabcolsep}{3pt} \renewcommand{\arraystretch}{1.35}

	\begin{tabular}{@{}>{\raggedright\arraybackslash}p{5.0cm} c c c c c c@{}}
		\toprule \makecell{\textbf{Work}}
		                                                     & \makecell{$\boldsymbol{\mathcal{F}_{\mathrm{on}}}$} &
		\makecell{$\boldsymbol{\mathcal{F}_{\mathrm{off}}}$} & \makecell{\textbf{All
		three}                                                                                                       \\\textbf{$^{14}$N HFS}} & \makecell{\textbf{Three-}
		\\\textbf{axis}\\\textbf{readout}} & \makecell{\textbf{Zeeman-pair}			    \\\textbf{$T$ comp.}}
		                                                     & \makecell{\textbf{Bias-field}                         \\\textbf{corr.}}
		\\ \midrule

		\mbox{\textbf{This work}}
		                                                     &
		\(\mathbf{1.2\pm0.1}\)
		                                                     &
		\(\mathbf{1.0\pm0.1}\)
		                                                     &
		\textbf{No}
		                                                     &
		\textbf{Yes}
		                                                     &
		\textbf{Yes}
		                                                     &
		\textbf{Yes}                                                                                                 \\

		\mbox{N. Sekiguchi \textit{et al.} (2024)~\cite{Sekiguchi2024}}
		                                                     &
		1.5
		                                                     &
		1.3
		                                                     &
		Yes
		                                                     &
		No
		                                                     &
		No
		                                                     &
		No                                                                                                           \\

		\mbox{Y. Kainuma \textit{et al.} (2024)~\cite{Kainuma2024}}
		                                                     &
		\(1.6^{*}\)
		                                                     &
		\(1.4^{*}\)
		                                                     &
		Yes
		                                                     &
		No
		                                                     &
		No
		                                                     &
		No                                                                                                           \\

		\mbox{S. M. Graham \textit{et al.} (2023)~\cite{Graham2023}}
		                                                     &
		2.7
		                                                     &
		1.9
		                                                     &
		Yes
		                                                     &
		No
		                                                     &
		No
		                                                     &
		No                                                                                                           \\

		\mbox{R. L. Patel \textit{et al.} (2020)~\cite{Patel2020}}
		                                                     &
		2.4
		                                                     &
		1.9
		                                                     &
		Yes
		                                                     &
		No
		                                                     &
		No
		                                                     &
		No                                                                                                           \\

		\mbox{Y. Wang \textit{et al.} (2025)~\cite{Wang2025Integrated}}
		                                                     &
		3.1
		                                                     &
		2.1
		                                                     &
		Yes
		                                                     &
		No
		                                                     &
		No
		                                                     &
		No                                                                                                           \\

		\mbox{J. H. Shim \textit{et al.} (2022)~\cite{Shim2022}}
		                                                     &
		\(5.2^{*}\)
		                                                     &
		\(3.0^{*}\)
		                                                     &
		No
		                                                     &
		No
		                                                     &
		Yes
		                                                     &
		No                                                                                                           \\

		\mbox{J. M. Schloss \textit{et al.} (2018)~\cite{Schloss2018}}
		                                                     &
		\(3.0\pm0.4^{*}\)
		                                                     &
		NR
		                                                     &
		Yes
		                                                     &
		Yes
		                                                     &
		No
		                                                     &
		No                                                                                                           \\

		\mbox{F. M. St\"urner \textit{et al.} (2021)~\cite{Sturner2021}}
		                                                     &
		\(13.2^{*}\)
		                                                     &
		\(14.0^{*}\)
		                                                     &
		Yes
		                                                     &
		No
		                                                     &
		No
		                                                     &
		No                                                                                                           \\

		\mbox{N. Wang \textit{et al.} (2026)~\cite{Wang2026ClosedLoop}}
		                                                     &
		NR
		                                                     &
		NR
		                                                     &
		Yes
		                                                     &
		No
		                                                     &
		No
		                                                     &
		No                                                                                                           \\

		\bottomrule
	\end{tabular}

	\caption{\raggedright\textbf{Benchmark against representative high-sensitivity ensemble
		NV magnetometers.} The noise factor is defined as \(\mathcal{F}=S/S_{\mathrm{limit}}\), where
	\(S_{\mathrm{limit}}=\sqrt{S_{\mathrm{elec}}^{2}+S_{\mathrm{PSN}}^{2}}\) is the readout limit set by the
	electronic and photon shot-noise contributions.  \(\mathcal{F}_{\mathrm{on}}\) denotes active ODMR tracking
	and \(\mathcal{F}_{\mathrm{off}}\) the corresponding magnetically insensitive off-resonance condition;
	hence, \(\mathcal{F}=1\) denotes operation at the calculated fundamental readout limit.  Values marked
	\(^{*}\) are normalized to the reported photon shot-noise floor where the electronic contribution could
	not be reconstructed and should therefore be interpreted as proxy noise factors. HFS denotes hyperfine
	structure and NR denotes not reported.}

	\label{tab:benchmark}
\end{table}

The comparison in Table~\ref{tab:benchmark} places the presented results close to this limit. During on-resonance operation, the sensor achieves \(\mathcal{F}_{\mathrm{on}}=1.2\pm0.1\), while simultaneously tracking six ODMR transitions for three-axis reconstruction. Under magnetically insensitive off-resonance operation, we obtain \(\mathcal{F}_{\mathrm{off}}=1.0\pm0.1\), showing that the measured baseline is experimentally indistinguishable from the calculated noise floor due to electronic noise and photon shot-noise. These results are the closest to the fundamental off-resonance limit among the systems reported in literature in Table~\ref{tab:benchmark}; with previously reported balanced and digitally corrected architectures possessing noise factors measurably above unity. 

More importantly, the distinction is not limited to noise performance. The sensor system reported in this work is the only architecture in the comparison that combines near-noise limit readout with simultaneous three-axis reconstruction, Zeeman-split pair rejection of temperature-induced zero-field-splitting shifts, and elimination of temperature-dependent bias-field drift within the same continuously operating measurement chain. By integrating these functions into a single, continuously operating measurement framework, this platform is explicitly designed for high-fidelity three-axis quantum magnetometry directly in unshielded field environments.

\section{Conclusion} \label{sec:conclusions} In summary, we have established a hardware-independent quantum processing architecture that systematically eliminates technical noise and long-term environmental drift entirely within the digital domain.  By integrating a time-adaptive digital photodetector differential scheme with real-time second-derivative ODMR slope drift correction feedback, this framework suppresses high-frequency correlated laser-intensity noise and dynamic gain variations during active operation.  This multi-stage digital stabilization results in the measured off-resonance noise to be indistinguishable from 
the calculated noise floor limit, marking a critical milestone toward reaching true photon shot-noise-limited operation in unshielded environments. Building upon this stabilized readout, our magnetic eigenvector transformation formalism effectively decouples macro-environmental thermal variations from the pure magnetic signal component without requiring additional complex physical infrastructure.  Ultimately, resolving the conflict between short-term precision and multi-hour baseline drift represents a significant paradigm shift in quantum instrumentation, providing a scalable blueprint for deploying a new generation of truly field-ready quantum sensors.

\section*{Methods}

\subsection*{Experimental system and three-axis readout} \label{sec:methods_system}

The sensor was built using a \(3\times3\times0.5~\mathrm{mm}^{3}\), \(^{12}\mathrm{C}\)-enriched ensemble-NV diamond with \(\{100\}\) major faces. The diamond was mounted directly above the central region of a custom \(\Omega\)-shaped microwave antenna fabricated on Rogers RO4350B substrate. Optical excitation was provided by a fiber-coupled \(520~\mathrm{nm}\) laser and focused onto the diamond through an aspheric lens. Unless stated otherwise, sensitivity and noise-floor measurements were obtained for an incident optical power of approximately \(363~\mathrm{mW}\); the power-dependent measurements were acquired over the power range shown in Fig.~\ref{fig:adaptive_readout}(d). A static bias field of approximately \(1~\mathrm{mT}\), produced by symmetrically positioned permanent magnets, separated ODMR transitions associated with the different NV orientations.

The fluorescence signal and the reference laser light were collected through the same sensor-head optical path and separated only after the diamond. A long-pass dichroic routed the red-shifted NV fluorescence to PD2, while the residual green excitation was directed to PD1 and used as the optical reference. Additional spectral filtering suppressed pump leakage into the fluorescence chain and fluorescence leakage into the reference chain. The two photodetector outputs were converted to voltage signals independently by transimpedance amplifiers and digitised without analog subtraction; differential cancellation was performed subsequently post digitisation.

Microwave excitation and synchronous detection were implemented on an AMD Zynq$-7020$ system-on-chip platform with two \(125~\mathrm{MS\,s^{-1}}\), $14-$bit ADC and DAC channels. Six independently programmable frequency-modulated microwave carriers addressed the the symmetric pairs of monitored spin transitions across three distinct NV crystallographic orientations.
The carriers were synthesized as a summed complex-baseband waveform, converted through the two DAC channels, and upconverted using an external IQ mixer driven by a common local oscillator near \(2.87~\mathrm{GHz}\).  The resulting multi-tone signal was amplified through a common microwave chain and delivered to the \(\Omega\)-antenna. The fluorescence and reference detector streams were sampled simultaneously and, for each carrier, demodulated at both the fundamental modulation frequency and its second harmonic. The resulting \(1f\) and \(2f\) channels were low-pass filtered using eighth-order filters with a cut-off frequency of approximately \(102~\mathrm{Hz}\) and retained at \(5~\mathrm{kS\,s^{-1}}\). The modulation frequencies were placed on a half-integer frequency comb: 

\begin{equation}
	f_{m,k} = \left(n+\frac{1}{2}+k\right)D, \qquad k=0,\ldots,5, \label{eq:modulation_comb}
\end{equation} with \(D=297.15~\mathrm{Hz}\). This places the nearest common inter-channel products at \(D/2=148.575~\mathrm{Hz}\) and \(D=297.15~\mathrm{Hz}\), well above the DC--\(100~\mathrm{Hz}\) magnetic readout band, where they are rejected digitally.

For NV orientation \(i\), the frequencies of the $m_s = 0 \rightarrow \pm1$ spin transitions are given by:
\begin{equation}
	\nu_{i,\pm} = D_{\mathrm{ZFS}}(T)\pm\gamma_{\mathrm e}B_i , \label{eq:zeeman_pair_methods}
\end{equation} 
where \(D_{\mathrm{ZFS}}(T)\) is the temperature-dependent zero-field splitting parameter, and \(B_i\) is the magnetic-field projection along the corresponding NV axis. The three simultaneously measured magnetic coordinates are therefore expressed as:

\begin{equation}
	B_i = \frac{\nu_{i,+}-\nu_{i,-}} {2\gamma_{\mathrm e}}, \qquad i\in\{1,2,3\}, \label{eq:three_axis_readout}
\end{equation} 
while the corresponding common-mode coordinate (representing the thermal coordinate),
\begin{equation}
	C_i = \frac{\nu_{i,+}+\nu_{i,-}}{2}, \label{eq:common_mode_coordinate}
\end{equation} 
isolates frequency shifts common to both spin transitions. 

Thus, \(B_1\), \(B_2\), and \(B_3\) provide the simultaneous three-axis magnetic readout, whereas \(C_i\) provides the thermal coordinate.

\subsection*{Noise cancellation and slope-normalized resonance tracking} \label{sec:methods_tracking}

The fluorescence and optical-reference channels contain strongly correlated laser-intensity noise, but
their relative gain and phase delay are not identical and can drift slowly during long-term
measurements. Instead of using a fixed subtraction scheme, the reference was matched to the fluorescence channel digitally after lock-in demodulation, following the general strategy of active optical-noise suppression used in ensemble-NV magnetometers \cite{Wang2025Laser,Wang2026ClosedLoop}. For each transition \(j\), independent filters were applied to the first- and second-derivative channels. Denoting either demodulated fluorescence channel by \(y_j[n]\) and the corresponding reference by \(r_j[n]\), the cleaned output is expressed as: 

\begin{equation}
	\widetilde{y}_j[n-4] = y_j[n-4] - \sum_{k=0}^{8} c_{j,k} r_j[n-k].  \label{eq:adaptive_cleaner}
\end{equation} 

The centred nine-tap FIR operates at \(5~\mathrm{kS\,s^{-1}}\), with a relative delay of up to \(\pm0.8~\mathrm{ms}\). Coefficients were initialized from an off-resonance record using ridge-regularized least squares and subsequently updated by bounded normalized least-mean-square adaptation~\cite{Slock1993}. Separate coefficient sets were maintained for the \(1f\) and \(2f\) outputs of all six transitions. The adaptation was constrained around the calibrated solution and kept slow
compared with the magnetic readout so that it followed gradual changes in the
relative response of the two optical channels rather than the ODMR signal. Long-duration cancellation and preservation of the ODMR lineshape are shown in Supplementary Information.

For a frequency-modulated ODMR transition with peak deviation \(\Delta\nu_j\), the demodulated derivative channels satisfy, to lowest order, \begin{equation}
	V_{1f,j}\propto \Delta\nu_j V_j', \qquad V_{2f,j}\propto \Delta\nu_j^2 V_j'', \label{eq:derivative_relation}
\end{equation} so that the \(1f\) resonant point provides the signed detuning while the \(2f\) response tracks the
local curvature~\cite{Wojciechowski2019}. Near the resonance frequency, this curvature is proportional to the first-derivative discriminator slope. Because finite modulation depths are used experimentally, the proportionality was calibrated from the measured ODMR response rather than inferred from the small-modulation approximation. If \(e_j[n]\) is the cleaned \(1f\) error and \(q_j[n]\) the cleaned \(2f\) signal, the live discriminator slope was evaluated as:

\begin{equation}
	K_j[n] = \kappa_j\,\mathcal{L}_{2f}\!\left\{q_j[n]\right\}, \label{eq:live_discriminator}
\end{equation} and the residual frequency detuning as:

\begin{equation}
	\Delta\nu_j^{\mathrm{res}}[n] = - \frac{e_j[n]} {\operatorname{sgn}\!\left(K_j[n]\right)
	\max\!\left(|K_j[n]|,K_{\min,j}\right)}.  \label{eq:slope_normalized_detuning}
\end{equation} 

Here, \(\mathcal{L}_{2f}\{\cdot\}\) denotes the low-pass filtering operation applied to the \(2f\) channel, and \(K_{\min,j}\) is the lower bound imposed on the discriminator slope's magnitude during normalization. These definitions prevent rapid gain fluctuations and unstable normalization when the discriminator slope becomes too small.

Each normalized detuning drove an independent digital \(\alpha\)--\(\beta\) tracker with an approximately \(8~\mathrm{Hz}\) closed-loop bandwidth. The tracker maintains estimates of the resonance point and its drift rate, keeping the microwave carrier near resonance while faster variations remain in the residual detuning; the complete measured resonance frequency is therefore:

\begin{equation}
	\nu_j^{\mathrm{read}}[n] = \nu_j^{\mathrm{prog}} + \Delta\nu_j^{\mathrm{ctrl}}[n] +
	\Delta\nu_j^{\mathrm{res}}[n].	\label{eq:frequency_readout}
\end{equation} 

This controller-plus-residual construction preserves the DC--\(100~\mathrm{Hz}\) magnetic readout
bandwidth despite the slower carrier-tracking loop~\cite{Lei2025,Wang2023HDR}. Extended closed-loop validation is
provided in Supplementary Information.
Before performing on-resonance measurements, an off-resonance record was used to initialize the adaptive filter, after which the six resonances were located and the demodulation phase, frequency deviation, microwave power,  resonance frequency, and local discriminator slope were calibrated independently for each transition. Opposite outermost \(^{14}\mathrm{N}\) hyperfine components were selected on the two Zeeman splits to avoid the reduction in first-derivative discriminator slope observed under simultaneous excitation of the same nuclear-spin manifold~\cite{Shim2022}. The corresponding phase, drive-parameter, and resonance-frequency calibrations are shown in Supplementary Information.

\subsection*{Magnetic eigenvector transformation and multiscale drift correction}
\label{sec:methods_eigen}

\noindent\textbf{Ambient-temperature-driven drift model.}
The long-term correction is formulated using the two Zeeman splits of a
single NV orientation. To first order, their resonance frequencies are
described by~\cite{Clevenson2018}
\begin{equation}
    f_{0,\pm}
    \approx
    D_{\mathrm{ZFS}}
    +
    \beta_T \Delta T
    \pm
    \gamma_{\mathrm e} B_{\mathrm{ext}},
    \label{eq:met_static_model}
\end{equation}
where \(D_{\mathrm{ZFS}}\approx2.87~\mathrm{GHz}\) is the zero-field
splitting, \(\beta_T\approx-74~\mathrm{kHz\,K^{-1}}\) is its temperature
coefficient near room temperature, \(\gamma_{\mathrm e}\) is the electron
gyromagnetic ratio, and \(B_{\mathrm{ext}}\) is the magnetic-field projection
along the selected NV axis. The temperature dependence may equivalently be
absorbed into the zero-field splitting,
\begin{equation}
    f_{0,\pm}
    \approx
    D_{\mathrm{ZFS}}(T)
    \pm
    \gamma_{\mathrm e} B_{\mathrm{ext}}.
    \label{eq:met_static_model_temperature}
\end{equation}

A change in \(D_{\mathrm{ZFS}}(T)\) produces a common shift of the two Zeeman splits, whereas a magnetic-field change produces equal and opposite frequency shifts. During long measurements, ambient temperature also changes the
field produced by the permanent bias magnets. The two tracked resonance
frequencies therefore contain both the common thermal shift of the NV
resonances and the temperature-dependent change in the bias magnetic field.

We write the corresponding time-dependent frequency deviations as
\begin{align}
    f_1(t)
    &=
    a(t)+b(t)+x(t)+y(t),
    \nonumber\\
    f_2(t)
    &=
    b(t)-a(t)+y(t)-x(t),
    \label{eq:met_dynamic_model}
\end{align}
with all terms expressed in frequency units. Here,
\(x(t)=\gamma_{\mathrm e}B_{\mathrm{ext}}(t)\) denotes the external magnetic
signal, \(a(t)\) the temperature-dependent drift of the bias magnetic field,
\(b(t)\) the slow common-mode thermal shift of the NV resonances, and \(y(t)\)
the faster common-mode contribution associated with optical and
laser-induced thermal fluctuations. Since \(a(t)\) and \(b(t)\) are driven by
the same ambient-temperature variation, they evolve in a correlated manner
over long measurement intervals.

The conventional Zeeman-split pair difference and pair mean are then
\begin{equation}
    \frac{f_1(t)-f_2(t)}{2}
    =
    a(t)+x(t),
    \qquad
    \frac{f_1(t)+f_2(t)}{2}
    =
    b(t)+y(t).
    \label{eq:met_pair_coordinates}
\end{equation}
The differential coordinate contains the magnetic signal together with the
temperature-dependent bias-field drift, while the pair mean follows the
common-mode thermal and optical evolution of the two resonance frequencies.

\vspace{0.3cm}

\noindent\textbf{Magnetic eigenvector transformation.}
The two resonance-frequency traces are mean-centred over the analysis
interval,
\begin{equation}
    \widetilde{\mathbf f}(t)
    =
    \begin{bmatrix}
        \widetilde f_1(t)\\
        \widetilde f_2(t)
    \end{bmatrix}
    =
    \begin{bmatrix}
        f_1(t)-\langle f_1\rangle\\
        f_2(t)-\langle f_2\rangle
    \end{bmatrix},
    \label{eq:met_mean_centering}
\end{equation}
and their empirical covariance matrix is
\begin{equation}
    \mathbf{\Sigma}
    =
    \left\langle
        \widetilde{\mathbf f}(t)
        \widetilde{\mathbf f}^{\mathsf T}(t)
    \right\rangle.
    \label{eq:met_covariance}
\end{equation}
The principal directions of the two-frequency measurement space are obtained
from
\begin{equation}
    \mathbf{\Sigma}\mathbf v_k
    =
    \lambda_k \mathbf v_k.
    \label{eq:met_eigenproblem}
\end{equation}

The magnetic eigenvector is identified from the known symmetry of the Zeeman
response. A magnetic perturbation produces equal and opposite frequency shifts
of the two Zeeman splits and therefore defines the reference vector
\begin{equation}
    \mathbf{u}_B
    =
    \frac{1}{\sqrt{2}}
    \begin{bmatrix}
        1\\
        -1
    \end{bmatrix}.
    \label{eq:met_ideal_magnetic_axis}
\end{equation}
The normalized covariance eigenvector with the larger overlap with
\(\mathbf u_B\) is assigned as the magnetic eigenvector
\(\mathbf v_B\). Its sign is chosen so that
\begin{equation}
    \mathbf v_B^{\mathsf T}\mathbf u_B>0,
\end{equation}
which fixes the polarity to that of the conventional Zeeman-split pair
difference. The orthogonal eigenvector represents the second principal component of the resonance-frequency data and captures the temperature-correlated variation of the pair.

Writing
\begin{equation}
    \mathbf v_B
    =
    \begin{bmatrix}
        v_{B1}\\
        v_{B2}
    \end{bmatrix},
\end{equation}
the magnetic projection is
\begin{align}
    p_B^{\mathrm{raw}}(t)
    &=
    \mathbf v_B^{\mathsf T}
    \widetilde{\mathbf f}(t)
    \nonumber\\
    &=
    (v_{B1}-v_{B2})
    \left[
        \widetilde a(t)+\widetilde x(t)
    \right]
    +
    (v_{B1}+v_{B2})
    \left[
        \widetilde b(t)+\widetilde y(t)
    \right].
    \label{eq:met_raw_projection}
\end{align}

The projected coordinate is normalized to preserve its magnetic-field
conversion factor,
\begin{equation}
    B_{\mathrm{eig}}^{\mathrm{raw}}(t)
    =
    \frac{
        \mathbf v_B^{\mathsf T}
        \widetilde{\mathbf f}(t)
    }{
        \sqrt{2}\,
        \gamma_{\mathrm e}
        \left(
            \mathbf v_B^{\mathsf T}\mathbf u_B
        \right)
    }.
    \label{eq:met_field_normalization}
\end{equation}
For a magnetic perturbation
\(\delta\mathbf f
=\sqrt{2}\gamma_{\mathrm e}\delta B\,\mathbf u_B\),
this normalization gives the corresponding field change directly. For
\(\mathbf v_B=\mathbf u_B\), it reduces to the usual Zeeman-split pair expression
\((f_1-f_2)/(2\gamma_{\mathrm e})\).

\vspace{0.3cm}

\noindent\textbf{Common-mode regression.}
The pair mean provides a convenient reference for the common-mode component
remaining in the magnetic eigenvector,
\begin{equation}
    C(t)
    =
    \frac{f_1(t)+f_2(t)}{2}
    =
    b(t)+y(t).
    \label{eq:met_common_mode}
\end{equation}
For the long-duration measurement, its slowly varying baseline is estimated
with a \(720~\mathrm{s}\) moving average. The faster component is expressed in
field-equivalent units as
\begin{equation}
    Y_{\mathrm{HF}}(t)
    =
    \frac{
        C(t)
        -
        \mathcal M_{720\mathrm{s}}
        \left\{
            C(t)
        \right\}
    }{
        \gamma_{\mathrm e}
    },
    \label{eq:met_hf_reference}
\end{equation}
where
\(\mathcal M_{720\mathrm{s}}\{\cdot\}\) denotes the
\(720~\mathrm{s}\) moving-average operation.

The component correlated with \(Y_{\mathrm{HF}}(t)\) is removed by linear
regression,
\begin{equation}
    B_1(t)
    =
    B_{\mathrm{eig}}^{\mathrm{raw}}(t)
    -
    \alpha Y_{\mathrm{HF}}(t),
    \label{eq:met_stage1}
\end{equation}
with
\begin{equation}
    \alpha
    =
    \frac{
        \operatorname{Cov}
        \left(
            B_{\mathrm{eig}}^{\mathrm{raw}},
            Y_{\mathrm{HF}}
        \right)
    }{
        \operatorname{Var}
        \left(
            Y_{\mathrm{HF}}
        \right)
    }.
    \label{eq:met_alpha}
\end{equation}

Slow changes in the optical excitation can remain after this step. Two
references were considered for their correction: an independent reference
photodetector and the second-harmonic ODMR response.

\vspace{0.2cm}

\noindent\textit{1. Reference-photodetector-based correction.}
An independent reference photodetector provides a direct measurement of
changes in the excitation intensity. Let \(I_{\mathrm{ref}}(t)\) denote its
output. After removing the mean, the slowly varying component is obtained by
low-pass filtering,
\begin{equation}
    I_{\mathrm{ref,LP}}(t)
    =
    \mathcal L_{\mathrm{LP}}
    \left\{
        I_{\mathrm{ref}}(t)
        -
        \left\langle I_{\mathrm{ref}}\right\rangle
    \right\},
    \label{eq:met_pd_reference}
\end{equation}
where \(\mathcal L_{\mathrm{LP}}\{\cdot\}\) denotes the low-pass filtering
operation.

The magnetic signal is then corrected as
\begin{equation}
    B_{\mathrm{corr}}^{(\mathrm{PD})}(t)
    =
    B_1(t)
    -
    \beta I_{\mathrm{ref,LP}}(t),
    \label{eq:met_pd_correction}
\end{equation}
where
\begin{equation}
    \beta
    =
    \frac{
        \operatorname{Cov}
        \left(
            B_1,
            I_{\mathrm{ref,LP}}
        \right)
    }{
        \operatorname{Var}
        \left(
            I_{\mathrm{ref,LP}}
        \right)
    }.
    \label{eq:met_beta}
\end{equation}

\vspace{0.2cm}

\noindent\textit{2. Second-harmonic-based correction.}
The second-harmonic (\(2f\)) ODMR response provides an intrinsic reference for
the same slow optical variation. At the resonance frequency, the \(2f\)
signal follows the local curvature of the ODMR line. Changes in optical power,
ODMR contrast, or linewidth therefore appear in the measured second-harmonic
response.

For the selected Zeeman pair, the calibrated \(2f\) responses are combined,
their baseline is removed, and the remaining slow component is isolated by
low-pass filtering to obtain \(S_{2f,\mathrm{LP}}(t)\). The corresponding
correction is
\begin{equation}
    B_{\mathrm{corr}}^{(2f)}(t)
    =
    B_1(t)
    -
    \kappa S_{2f,\mathrm{LP}}(t),
    \label{eq:met_2f_correction}
\end{equation}
with
\begin{equation}
    \kappa
    =
    \frac{
        \operatorname{Cov}
        \left(
            B_1,
            S_{2f,\mathrm{LP}}
        \right)
    }{
        \operatorname{Var}
        \left(
            S_{2f,\mathrm{LP}}
        \right)
    }.
    \label{eq:met_kappa}
\end{equation}

The two approaches differ only in the source of low-frequency optical
reference. The photodetector route uses a separately measured intensity
signal, whereas the second-harmonic route derives the reference from ODMR
response itself. In the measurements reported here, the second-harmonic route
was used for the final correction, since the \(2f\) channels were already
acquired continuously as part of the resonance-tracking architecture.

\subsection*{Fundamental readout-noise evaluation} \label{sec:methods_noise}

The fundamental readout floor was evaluated from the independently calculated photon- hot-noise and electronic noise contributions, with all terms referred to the differential Zeeman-split pair magnetic readout using the measured ODMR voltage-to-frequency discriminator slope. At the \(363~\mathrm{mW}\) operating point used for off-resonance calibration, the measured photocurrent was \(I_{\mathrm{dc}}=341~\mu\mathrm{A}\), the transimpedance gain was \(R_{\mathrm{TIA}}=18~\mathrm{k\Omega}\), and the first-derivative discriminator slope magnitude was \(|m|=29.8~\mathrm{mV\,MHz^{-1}}\). Using \(\gamma_{\mathrm e}=28~\mathrm{Hz\,nT^{-1}}\), the corresponding photon shot-noise contribution is:

\begin{equation}
	S_{\mathrm{PSN}} = \frac{
		R_{\mathrm{TIA}}\sqrt{2qI_{\mathrm{dc}}}
	}{
		|m|\gamma_{\mathrm e}
	} = 225.3~\mathrm{pT/\sqrt{Hz}}.  \label{eq:psn_limit}
\end{equation}

The electronic contribution includes contributions from the independent transimpedance-amplifier and ADC quantisation noise sources:

\begin{equation}
	S_{\mathrm{elec}} = \sqrt{
		S_{\mathrm{TIA}}^{2} + S_{\mathrm{ADC}}^{2}
	} = 174.1~\mathrm{pT/\sqrt{Hz}}.  \label{eq:electronic_limit}
\end{equation} The calculated off-resonance readout limit is therefore \begin{equation}
	S_{\mathrm{limit}} = \sqrt{
		S_{\mathrm{PSN}}^{2} + S_{\mathrm{elec}}^{2}
	} = 284.7~\mathrm{pT/\sqrt{Hz}}.  \label{eq:fundamental_readout_limit}
\end{equation} The experimentally measured electronic and off-resonance floors were \(177\pm4~\mathrm{pT/\sqrt{Hz}}\) and \(288\pm11~\mathrm{pT/\sqrt{Hz}}\), respectively. The latter corresponds to \(1.01\,S_{\mathrm{limit}}\), indicating that no resolvable laser-noise-dominated excess remains after adaptive cancellation within the measured band. Microwave phase and amplitude noise are relevant only during resonant operation and are not included in the off-resonance limit. Complete component-level derivations, including detector, TIA, ADC, microwave, and laser-intensity-noise contributions, are provided in Supplementary Information.

\vspace{0.6em}
\section*{Data availability}
The data that support the findings of this study are available from
the corresponding authors upon reasonable request.

\section*{Code availability}
The code used for data analysis in this study is available from the corresponding
authors upon reasonable request.

\printbibliography

@article{Webb2019,
  author = {Webb, J. L. and Clement, J. D. and Troise, L. and Ahmadi, S. and Johansen, G. J. and Huck, A. and Andersen, U. L.},
  title = {Nanotesla Sensitivity Magnetic Field Sensing Using a Compact Diamond Nitrogen-Vacancy Magnetometer},
  journal = {Applied Physics Letters},
  volume = {114},
  pages = {231103},
  year = {2019},
  doi = {10.1063/1.5095241}
}

@article{Patel2020,
  author = {Patel, R. L. and Zhou, L. Q. and Frangeskou, A. C. and Stimpson, G. A. and Breeze, B. G. and Nikitin, A. and Dale, M. W. and Nichols, E. C. and Thornley, W. and Green, B. L. and Newton, M. E. and Edmonds, A. M. and Markham, M. L. and Twitchen, D. J. and Morley, G. W.},
  title = {Subnanotesla Magnetometry with a Fiber-Coupled Diamond Sensor},
  journal = {Physical Review Applied},
  volume = {14},
  pages = {044058},
  year = {2020},
  doi = {10.1103/PhysRevApplied.14.044058}
}

@article{Graham2023,
  author = {Graham, Stuart M. and Rahman, A. T. M. A. and Munn, L. and Patel, R. L. and Newman, A. J. and Stephen, C. J. and Colston, G. and Nikitin, A. and Edmonds, A. M. and Twitchen, D. J. and Markham, M. L. and Morley, G. W.},
  title = {Fiber-Coupled Diamond Magnetometry with an Unshielded Sensitivity of 30 pT/$\sqrt{\mathrm{Hz}}$},
  journal = {Physical Review Applied},
  volume = {19},
  pages = {044042},
  year = {2023},
  doi = {10.1103/PhysRevApplied.19.044042}
}

@article{Zhang2022Fiber,
  author = {Zhang, Shao-Chun and Lin, Hao-Bin and Dong, Yang and Du, Bo and Gao, Xue-Dong and Yu, Cui and Feng, Zhi-Hong and Chen, Xiang-Dong and Guo, Guang-Can and Sun, Fang-Wen},
  title = {High-Sensitivity and Wide-Bandwidth Fiber-Coupled Diamond Magnetometer with Surface Coating},
  journal = {Photonics Research},
  volume = {10},
  number = {9},
  pages = {2191--2201},
  year = {2022},
  doi = {10.1364/PRJ.462851}
}

@article{Kim2019,
  author = {Kim, Donggyu and Ibrahim, Mohamed I. and Foy, Christopher and Trusheim, Matthew E. and Han, Ruonan and Englund, Dirk R.},
  title = {A CMOS-Integrated Quantum Sensor Based on Nitrogen-Vacancy Centres},
  journal = {Nature Electronics},
  volume = {2},
  pages = {284--289},
  year = {2019},
  doi = {10.1038/s41928-019-0275-5}
}

@article{Sekiguchi2024,
  author = {Sekiguchi, Naota and Fushimi, M. and Yoshimura, A. and Shinei, C. and Miyakawa, M. and Taniguchi, T. and Teraji, T. and Abe, H. and Onoda, S. and Ohshima, T. and Hatano, M. and Sekino, M. and Iwasaki, T.},
  title = {Diamond Quantum Magnetometer with DC Sensitivity of Sub-10 pT/$\sqrt{\mathrm{Hz}}$ toward Measurement of Biomagnetic Field},
  journal = {Physical Review Applied},
  volume = {21},
  pages = {064010},
  year = {2024},
  doi = {10.1103/PhysRevApplied.21.064010}
}

@article{Kumar2024,
  author = {Kumar, H. and Dasika, S. and Mangat, M. and Tallur, S. and Saha, K.},
  title = {High Dynamic-Range and Portable Magnetometer Using Ensemble Nitrogen-Vacancy Centers in Diamond},
  journal = {Review of Scientific Instruments},
  volume = {95},
  number = {7},
  pages = {075002},
  year = {2024},
  doi = {10.1063/5.0205105}
}

@article{Huang2024,
  author = {Huang, Kun and Mao, Xiaobiao and Zhang, Yu and Wang, Mengzhu and He, Xinhui and Ran, Guihao and Hu, Qin and Lin, Zhennan},
  title = {A Portable and Highly Integrated Solid-State Quantum Magnetometer Module Based on the Diamond NV Color Centers},
  journal = {IEEE Transactions on Instrumentation and Measurement},
  volume = {73},
  pages = {9518009},
  year = {2024},
  doi = {10.1109/TIM.2024.3470972}
}

@article{Zheng2024,
  author = {Zheng, Doudou and Liu, Zhenhua and Fu, Jianghao and Tang, Jun and Guo, Hao and Wen, Huanfei and Li, Zhonghao and Ma, Zongmin and Liu, Jun},
  title = {An Integrated Nitrogen-Vacancy Magnetometer with Subnanotesla Sensitivity for Practical Applications},
  journal = {IEEE Sensors Journal},
  volume = {24},
  pages = {34198--34204},
  year = {2024},
  doi = {10.1109/JSEN.2024.3446867}
}

@article{Liu2024Fiber,
  author = {Liu, Yankang and Liu, Zhenhua and Li, Yang and Ma, Zongmin and Li, Zhonghao and Wen, Huanfei and Guo, Hao and Tang, Jun and Li, Yanjun and Liu, Jun},
  title = {The Fiber Self-Focusing Integrated Nitrogen Vacancy Magnetometer},
  journal = {IEEE Transactions on Instrumentation and Measurement},
  volume = {73},
  pages = {1--8},
  year = {2024},
  doi = {10.1109/TIM.2024.3396855}
}

@article{Wang2025Integrated,
  author = {Wang, Yifan and Zhang, Wenzhe and Chai, Haotian and Zhang, Zhenlin and Lin, Shaochun and Qin, Xi and Du, Jiangfeng},
  title = {Fully Integrated Quantum Magnetometer Based on Nitrogen-Vacancy Centers},
  journal = {Physical Review Applied},
  volume = {23},
  pages = {034008},
  year = {2025},
  doi = {10.1103/PhysRevApplied.23.034008}
}

@article{Liu2026Chip,
  author = {Liu, Yong and Zang, Han-Xiang and Ma, Meng-Qi and Jiang, Wang and Bai, Zhe and Gao, Xue-Dong and Dong, Yang and Shi, Hong-yan and Du, Bo and Chen, Xiang-Dong and Zhang, Shao-Chun and Sun, Fang-Wen},
  title = {Fiber-Integrated Chip-Scale Diamond Quantum Magnetometer},
  journal = {Diamond and Related Materials},
  volume = {166},
  pages = {113723},
  year = {2026},
  doi = {10.1016/j.diamond.2026.113723}
}

@inproceedings{Xie2024MEMS,
  author = {Xie, F. and Chen, Z. and Peng, X. and Liu, Q. and Li, L. and Wang, N. and Hu, Y. and Liu, Y. and Wang, L. and Chen, H. and Cheng, J. and Wu, Z.},
  title = {Miniaturized Diamond Quantum Magnetometer with Integrated Laser Source and All Electrical I/Os},
  booktitle = {2024 IEEE 37th International Conference on Micro Electro Mechanical Systems (MEMS)},
  pages = {577--580},
  year = {2024},
  doi = {10.1109/MEMS58180.2024.10439546}
}

@article{Schloss2018,
  author = {Schloss, Jennifer M. and Barry, John F. and Turner, Matthew J. and Walsworth, Ronald L.},
  title = {Simultaneous Broadband Vector Magnetometry Using Solid-State Spins},
  journal = {Physical Review Applied},
  volume = {10},
  pages = {034044},
  year = {2018},
  doi = {10.1103/PhysRevApplied.10.034044}
}

@article{Clevenson2018,
  author = {Clevenson, Hannah and Pham, Linh M. and Teale, Carson and Johnson, Kerry and Englund, Dirk and Braje, Danielle},
  title = {Robust High-Dynamic-Range Vector Magnetometry with Nitrogen-Vacancy Centers in Diamond},
  journal = {Applied Physics Letters},
  volume = {112},
  pages = {252406},
  year = {2018},
  doi = {10.1063/1.5034216}
}

@article{Sturner2021,
  author = {St{\"u}rner, Felix M. and Brenneis, Andreas and Buck, Thomas and Kassel, Julian and R{\"o}lver, Robert and Fuchs, Tino and Savitsky, Anton and Suter, Dieter and Grimmel, Jens and Hengesbach, Stefan and F{\"o}rtsch, Michael and Nakamura, Kazuo and Sumiya, Hitoshi and Onoda, Shinobu and Isoya, Junichi and Jelezko, Fedor},
  title = {Integrated and Portable Magnetometer Based on Nitrogen-Vacancy Ensembles in Diamond},
  journal = {Advanced Quantum Technologies},
  volume = {4},
  pages = {2000111},
  year = {2021},
  doi = {10.1002/qute.202000111}
}

@article{Wang2026ClosedLoop,
  author = {Wang, Nan and Xue, Wenli and Peng, Xiao and Zhu, Yaochen and Xu, Changbao and Hu, Yuqiang and Chen, Dalong and Su, Yongquan and Wang, Lihao and Liu, Yichen and Liu, Qihui and Wu, Zhenyu and Chen, Hao},
  title = {Closed-Loop Laser Noise Suppression with a Variable Optical Attenuator for Fiber-Integrated Diamond Quantum Sensor},
  journal = {Journal of Physics D: Applied Physics},
  volume = {59},
  number = {2},
  pages = {025010},
  year = {2026},
  doi = {10.1088/1361-6463/ae3329}
}

@article{Acosta2010,
  author = {Acosta, V. M. and Bauch, E. and Ledbetter, M. P. and Waxman, A. and Bouchard, L. S. and Budker, D.},
  title = {Temperature Dependence of the Nitrogen-Vacancy Magnetic Resonance in Diamond},
  journal = {Physical Review Letters},
  volume = {104},
  pages = {070801},
  year = {2010},
  doi = {10.1103/PhysRevLett.104.070801}
}

@article{Wang2025Laser,
  author = {Wang, Nan and Liu, Yichen and Su, Yongquan and Peng, Xiao and Hu, Yuqiang and Liu, Qihui and Xie, Fei and Zhu, Yaochen and Chen, Xin and Luo, Xin and Zhang, Yonggui and Wang, Lihao and Jing, Maoheng and Li, Chun and Nie, Shaoxiong and Chen, Hao and Wu, Zhenyu and Cheng, Jiangong},
  title = {Microfabricated Active Laser Noise Suppression Device for a High-Sensitivity Diamond Quantum Magnetometer},
  journal = {ACS Photonics},
  volume = {12},
  number = {2},
  pages = {828--838},
  year = {2025},
  doi = {10.1021/acsphotonics.4c01825}
}

@article{Liu2019SmCo,
  author = {Liu, Lei and Liu, Zhuang and Zhang, Xin and Zhang, Chaoyue and Li, Tianyi and Lee, Don and Yan, Aru},
  title = {2:17 Type SmCo Magnets with Low Temperature Coefficients of Remanence and Coercivity},
  journal = {Journal of Magnetism and Magnetic Materials},
  volume = {473},
  pages = {376--380},
  year = {2019},
  doi = {10.1016/j.jmmm.2018.10.113}
}

@article{Kainuma2024,
  author = {Kainuma, Yuta and Hatano, Yuji and Shibata, Takayuki and Sekiguchi, Naota and Nakazono, Akimichi and Kato, Hiromitsu and Onoda, Shinobu and Ohshima, Takeshi and Hatano, Mutsuko and Iwasaki, Takayuki},
  title = {Compact and Stable Diamond Quantum Sensors for Wide Applications},
  journal = {Advanced Quantum Technologies},
  volume = {7},
  pages = {2300456},
  year = {2024},
  doi = {10.1002/qute.202300456}
}

@article{Shim2022,
  author = {Shim, Jeong Hyun and Lee, Seong-Joo and Ghimire, Santosh and Hwang, Ju Il and Lee, Kwang-Geol and Kim, Kiwoong and Turner, Matthew J. and Hart, Connor A. and Walsworth, Ronald L. and Oh, Sangwon},
  title = {Multiplexed Sensing of Magnetic Field and Temperature in Real Time Using a Nitrogen-Vacancy Ensemble in Diamond},
  journal = {Physical Review Applied},
  volume = {17},
  pages = {014009},
  year = {2022},
  doi = {10.1103/PhysRevApplied.17.014009}
}

@misc{Halde2025Field,
  author = {Halde, Vincent and Bernard, Olivier and Brochu, Mathieu and Dufresne, Laurier and Fleury, Nicolas and Johnson, Kayla and Moffet, Benjamin and Roy-Guay, David},
  title = {Who Let the Diamonds Out?},
  year = {2025},
  eprint = {2509.19179},
  archivePrefix = {arXiv},
  primaryClass = {quant-ph},
  doi = {10.48550/arXiv.2509.19179},
  note = {arXiv preprint}
}

@misc{ConstantinidesRTC,
  author = {Constantinides, Steve},
  title = {Understanding and Using Reversible Temperature Coefficients},
  year = {2009},
  howpublished = {Arnold Magnetic Technologies technical publication},
  note = {Presented at MAGNETICS 2010},
  url = {https://www.arnoldmagnetics.com/wp-content/uploads/2017/10/Understanding-and-Using-Reversible-Temperature-Coefficients-Constantinides-Magnetics-2010-psn-hi-res.pdf}
}

@misc{EEC_SmCoTC,
  author = {{Electron Energy Corporation}},
  title = {Temperature Compensated Samarium Cobalt Magnets},
  howpublished = {Technical product data},
  note = {Accessed 19 August 2026},
  url = {https://www.electronenergy.com/temperature-compensated-samarium-cobalt-magnets/}
}

@article{Barry2020RMP,
  author = {Barry, John F. and Schloss, Jennifer M. and Bauch, Erik and Turner, Matthew J. and Hart, Connor A. and Pham, Linh M. and Walsworth, Ronald L.},
  title = {Sensitivity Optimization for {NV}-Diamond Magnetometry},
  journal = {Reviews of Modern Physics},
  volume = {92},
  pages = {015004},
  year = {2020},
  doi = {10.1103/RevModPhys.92.015004}
}

@article{Graham2025Road,
  author = {Graham, Stuart M. and Newman, Alex J. and Stephen, Colin J. and Edmonds, Andrew M. and Twitchen, Daniel J. and Markham, Matthew L. and Morley, Gavin W.},
  title = {On the Road with a Diamond Magnetometer},
  journal = {Diamond and Related Materials},
  volume = {152},
  pages = {111945},
  year = {2025},
  doi = {10.1016/j.diamond.2025.111945}
}

@article{Lei2025,
  author = {Lei, Yaowu and Zhou, Feng and Li, Dong and Chen, Dezhi and Hu, Haoliang and Li, Xiaofei and Ouyang, Zihao and Li, Yuhao and Zuo, Chen and Huang, Junchang},
  title = {Sensitivity Optimization for Lock-In Detection in Nitrogen-Vacancy Center-Based Magnetic Sensing Using Resonance Frequency Tracking Method},
  journal = {Diamond and Related Materials},
  volume = {159},
  pages = {112832},
  year = {2025},
  doi = {10.1016/j.diamond.2025.112832}
}

@article{Wang2023HDR,
  author = {Wang, Cao and Liu, Qihui and Hu, Yuqiang and Xie, Fei and Krishna, Krishangi and Wang, Nan and Wang, Lihao and Wang, Yang and Toussaint, Kimani C. and Cheng, Jiangong and Chen, Hao and Wu, Zhenyu},
  title = {Realization of High-Dynamic-Range Broadband Magnetic-Field Sensing with Ensemble Nitrogen-Vacancy Centers in Diamond},
  journal = {Review of Scientific Instruments},
  volume = {94},
  pages = {015109},
  year = {2023},
  doi = {10.1063/5.0089908}
}

@article{Zhao2019,
  author = {Zhao, Binbin and Guo, Hao and Zhao, Rui and Du, Fangfang and Li, Zhonghao and Wang, Lei and Wu, Dajin and Chen, Yulei and Tang, Jun and Liu, Jun},
  title = {High-Sensitivity Three-Axis Vector Magnetometry Using Electron Spin Ensembles in Single-Crystal Diamond},
  journal = {IEEE Magnetics Letters},
  volume = {10},
  pages = {8101104},
  year = {2019},
  doi = {10.1109/LMAG.2019.2891616}
}

@article{Childress2025BiasFree,
  author = {Childress, Lilian and Halde, Vincent and Johnson, Kayla and Lowther, Andrew and Roy-Guay, David and Ruhlmann, Romain and Solyom, Adrian},
  title = {Bias-Field-Free Operation of Nitrogen-Vacancy Ensembles in Diamond for Accurate Vector Magnetometry},
  journal = {PRX Quantum},
  volume = {6},
  pages = {040364},
  year = {2025},
  doi = {10.1103/zcdm-5qq3}
}

@article{Wojciechowski2019,
  author = {Wojciechowski, Adam M. and Nakonieczna, Paulina and Mr{\'o}zek, Mariusz and Sycz, Krystian and Ficek, Mateusz and G{\l}owacki, Maciej and Bogdanowicz, Robert and Gawlik, Wojciech},
  title = {Optical Magnetometry Based on Nanodiamonds with Nitrogen-Vacancy Color Centers},
  journal = {Materials},
  volume = {12},
  number = {18},
  pages = {2951},
  year = {2019},
  doi = {10.3390/ma12182951}
}

@article{Xie2023DoubleTransition,
  author = {Xie, Caijin and Zhu, Yunbin and Xie, Yijin and Li, Tingwei and Zhang, Wenzhe and Wang, Yifan and Rong, Xing},
  title = {Temperature-Robust Diamond Magnetometry Based on the Double-Transition Method},
  journal = {JUSTC},
  volume = {53},
  number = {7},
  pages = {0701},
  year = {2023},
  doi = {10.52396/JUSTC-2022-0150}
}

@article{Slock1993,
  author  = {Slock, Dirk T. M.},
  title   = {On the Convergence Behavior of the LMS and the Normalized LMS Algorithms},
  journal = {IEEE Transactions on Signal Processing},
  volume  = {41},
  number  = {9},
  pages   = {2811--2825},
  year    = {1993},
  doi     = {10.1109/78.236504}
}

\section*{Acknowledgments}
This work was supported by the Indian National Quantum Mission (NQM), Department of Science and Technology (DST), Government of India. The authors gratefully acknowledge the Indian Institute of Geomagnetism (IIG), Alibag, for providing access to observatory facilities. We thank Prof. Veeresh Deshpande and Prof. Rahul Singh at Indian Institute of Technology (IIT) Bombay, Prof. Saikat Ghosh at Indian Institute of Technology (IIT) Kanpur and Prof. Vidya Praveen Bhallamudi at Indian Institute of Technology (IIT) Madras for insightful discussions. We acknowledge  valuable technical support and assistance from Mr. Abhijeet Ghodgaonkar. The authors further acknowledge the support of the staff and access to facilities at the Wadhwani Electronics Lab (WEL), Department of Electrical Engineering, Indian Institute of Technology (IIT) Bombay.

\section*{Author information}

\subsection*{Authors and Affiliations}
\begin{enumerate}
	\item \textbf{Department of Electrical Engineering, Indian Institute of Technology Bombay, Mumbai, Maharashtra, India}\\
	      Annirudh K P, Shradha Atakar, Sanika Joshi, Maheshwar Mangat, Siddharth Tallur and Kasturi Saha
    \item \textbf{Qmet Tech Foundation, Mumbai, Maharashtra, India}\\
	      Annirudh K P, Vinayak Rane, Jay Gharat and Kasturi Saha
\end{enumerate}

\subsection*{Author contributions} 
AKP, VR, ST, and KS conceived the conceptual framework of the work. AKP and VR led the implementation and experimental execution of the proposed ideas, in collaboration with SSA on optical design. SJ, JG, and MM contributed to the experimental development, including noise modelling, electronics, and the power console. AKP, VR, and SSA prepared the manuscript, with contributions and feedback from all authors. ST and KS supervised the overall project.

\subsection*{Corresponding authors} Correspondence may be addressed to Kasturi Saha or Siddharth Tallur.

\section*{Competing interests} 
The work presented in this paper has been provisionally filed with Indian patent application number 202621109142. KS is also the project director of the Indian National Quantum Mission's hub for quantum sensing and metrology funded by the Department of Science and Technology (DST).

\clearpage
\end{document}